\documentclass{ptptex}
\usepackage{wrapft}
\usepackage{graphicx}

\newcommand{\kket}[1]{| {#1} \rangle\!\rangle}     %%
\newcommand{\wtilde}[1]{\widetilde{#1}} %%

\def\bsub{\begin{subequations}}
\def\esub{\end{subequations}}
\def\beq{\begin{eqnarray}}
\def\eeq{\end{eqnarray}}
\def\bsub{\begin{subequations}}
\def\esub{\end{subequations}}
\def\b{\begin{equation}}
\def\bs{\begin{split}}
\def\es{\end{split}}
\def\e{\end{equation}}
\begin{document}

\title{A practical scheme for constructing the minimum weight states of the 
$su(n)$-Lipkin model in arbitrary fermion number}

\author{Yasuhiko {\sc Tsue}$^{1,2}$, {Constan\c{c}a {\sc Provid\^encia}}$^{1}$, {Jo\~ao da {\sc Provid\^encia}}$^{1}$ and {Masatoshi {\sc Yamamura}}$^{1,3}$
%Thanks{These authors contributed equally to this work}}
}
%%%%%%%%%%% The \name command should be used as \name{Insert author name here}{Insert affiliation number here}
%%%%% Please use \thanks for contributed author details

%%%%%%%%%%% The \affil command should be used as \affil{Insert affiliation number here}{Insert author address here}
\inst{$^{1}${CFisUC, Departamento de F\'{i}sica, Universidade de Coimbra, 3004-516 Coimbra, 
Portugal}\\
$^{2}${Department of Mathematics and Physics, Kochi University, Kochi 780-8520, Japan}\\
%\affil{2}{Departamento de F\'{i}sica, Universidade de Coimbra, 3004-516 Coimbra, 
%Portugal}\\
$^{3}${Department of Pure and Applied Physics, 
Faculty of Engineering Science,\\
Kansai University, Suita 564-8680, Japan}\\
}

\abst{
With the aim of performing an argument supplement to the previous paper by the present authors, 
in this paper, a practical scheme for constructing the minimum weight states of the $su(n)$-Lipkin model in 
arbitrary fermion number is discussed. 
The idea comes from the following two points : (i) consideration 
on the property of one-fermion transfer induced by the $su(n)$-generators in the Lipkin model and 
(ii) use of the auxiliary $su(2)$-algebra presented by the present authors. 
The form obtained under the points (i) and (ii) is simple. 
}

%\subjectindex{xxxx, xxx}

\maketitle

As is well known, the Lipkin model proposed by Lipkin, Meshkov and Glick in 1965 \cite{1} is a classical one. 
This model is based on the $su(2)$-algebra and has played, in certain sense, a central role in the studies of nuclear many-body theories. 
As one of the theoretical interests, the original Lipkin model has been generalized to the form 
based on the $su(n)$-algebra. 
We call it as the $su(n)$-Lipkin model \cite{2}. 
However, the generalization has been restricted to the case of the ``closed-shell" system : 
In the absence of interaction, all fermions fully occupy energetically lowest single-particle level.

Under the above-mentioned situation, the present authors, recently, published three papers, in which the $su(n)$-Lipkin model was discussed for the 
case in arbitrary fermion number \citen{3}, \citen{4} and \citen{5}.
Particularly, in \citen{3} which will be referred to as (I), we discussed the minimum weight states in arbitrary fermion number. 
In the ``closed-shell" system, the minimum weight state is simply and uniquely given without any comment. 
However, the algebraic approach to many-body theories starts in the task 
how to express the minimum weight states. 
In the case with the ``closed-shell" system, we can skip this task. 
After finishing this step, 
we must construct the orthogonal set by 
operating the ``raising" operators on the minimum weight states appropriately. 
This problem was discussed in the frame of our idea \citen{4} and \citen{5}. 

In (I), we gave a possible idea for constructing the minimum weight states in the case with arbitrary fermion number. 
The basic idea can be found in the paper by the present authors \cite{6}~: 
Introduction of new $su(2)$-algebra into the $su(2)$-Lipkin model. 
Any of the three generators commutes with the three generators in the Lipkin model. 
We called it as the auxiliary $su(2)$-algebra. 
With the help of this algebra, we can determine the minimum weight states of the $su(2)$-Lipkin model. 
In this connection, the magnitude of this new $su(2)$-algebra in the case of the ``closed-shell" system is 
equal to zero. 
In (I), we generalize this case to the $su(n)$-Lipkin model. 
Naturally, in this case, we can also present the auxiliary $su(2)$-algebra generalized from the form in the $su(2)$-Lipkin model. 
The explicit form of the minimum weight states of the $su(n)$-Lipkin model is shown in the relation (I.5.15). 
As can be seen in this relation, the minimum weight states are expressed in terms of the raising operators of the auxiliary $su(2)$-algebras 
of the $su(n')$-Lipkin model for $n' \leq n$. 
However, the commutation relations among the generators with $n'$ and $n''\ (n'\neq n'')$ are complicated. 
Therefore, it may be very tedious to obtain the normalized states. 
The above tells us that the form (I.5.15) may be not practical. 
The aim of this paper is to give an alternative form of the minimum weight states, which may be expected to overcome the above-mentioned trouble 
to the normalization.

First, we recapitulate briefly this model in the form suitable for the present discussion. 
It consists of $(M+1)$ single-particle levels $(M=1,\ 2,\cdots )$, which are labeled 
as $i=0,\ 1,\ 2,\cdots\ ,\ M$. 
The case with $M=1$ corresponds to the original $su(2)$-Lipkin model \cite{1}. 
Each single-particle level contains $2\Omega$ single-particle states which are discriminated from each other by $\mu=1,\ 2,\cdots ,\ 2\Omega$. 
Then, every single-particle state can be specified by $(i,\ \mu)$. 
With the use of the fermion operators $({\tilde c}_{i,\mu},\ {\tilde c}_{i,\mu}^*)$, we define the following bi-linear form:
\beq\label{1}
& &{\wtilde S}^i=\sum_{\mu=1}^{2\Omega}{\tilde c}_{i,\mu}^*{\tilde c}_{0,\mu}\ , \qquad
{\wtilde S}_i=\sum_{\mu=1}^{2\Omega}{\tilde c}_{0,\mu}^*{\tilde c}_{i,\mu}\ , \qquad
(i=1,\ 2,\cdots ,\ M)\nonumber\\
& &{\wtilde S}_j^i=\sum_{\mu=1}^{2\Omega}\left({\tilde c}_{i,\mu}^*{\tilde c}_{j,\mu}-\delta_{ij}{\tilde c}_{0,\mu}^*{\tilde c}_{0,\mu}\right)\ . \qquad
(i,\ j=1,\ 2,\cdots ,\ M)
\eeq
The set $({\wtilde S}^i,\ {\wtilde S}_i,\ {\wtilde S}_j^i)$ forms the $su(M+1)$-algebra:
\bsub\label{2}
\beq
& &\left[\ {\wtilde S}^i\ , \ {\wtilde S}_j\ \right]={\wtilde S}_j^i\ , 
\label{2a}\\
& &\left[\ {\wtilde S}_j^i\ , \ {\wtilde S}^k\ \right]=\delta_{jk}{\wtilde S}^i+\delta_{ij}{\wtilde S}^k\ , 
\label{2b}\\
& &\left[\ {\wtilde S}_j^i\ , \ {\wtilde S}_k^l\ \right]=\delta_{jl}{\wtilde S}_k^i-\delta_{ik}{\wtilde S}_j^l\ , 
\label{2c}
\eeq
\esub
The $su(n)$-generators (\ref{1}) are expressed in the form 
$\sum_{\mu=1}^{2\Omega}{\tilde c}_{i,\mu}^*{\tilde c}_{j,\mu}$ for 
$i\neq j$ including the case with $i$ or $j=0$. 
In this form, we can see that $j$ changes to $i$, but $\mu$ does not change. 
Later, this property will play a key role for constructing the minimum weight states.

In association with the above $su(M+1)$-algebra, we can introduce new $su(2)$-algebra in the present fermion space. 
Two cases with $M=1$ and arbitrary value of $M$ have been discussed in \citen{6} and \citen{3}, respectively. 
We introduce the following operators:
\beq\label{3}
{\tilde d}_{\mu}^*=
\left\{
\begin{array}{ll}
\displaystyle \prod_{i=0}^M{\tilde c}_{i,\mu}^*\ , & (M=1,\ 3,\ 5,\cdots)\\
\displaystyle e_\mu\prod_{i=0}^M{\tilde c}_{i,\mu}^*\ , & (M=2,\ 4,\ 6,\cdots)\\
\end{array}
\right.
\eeq
Here, $e_{\mu}$ denotes the Clifford number obeying 
\beq\label{4}
e_\mu\cdot e_{\mu'}+e_{\mu'}\cdot e_{\mu}=2\delta_{\mu\mu'}\ . 
\eeq
With the use of the operator (\ref{3}), we define ${\wtilde \Lambda}_{\pm}$ in the form 
\beq\label{5}
{\wtilde \Lambda}_+=\sum_{\mu=1}^{2\Omega}{\tilde d}_{\mu}^*\ , \qquad
{\wtilde \Lambda}_-=\sum_{\mu=1}^{2\Omega}{\tilde d}_{\mu}\ .
\eeq
The operator ${\wtilde \Lambda}_0$ is defined as 
\beq\label{6}
{\wtilde \Lambda}_0=\frac{1}{2}\sum_{\mu=1}^{2\Omega}[\ {\tilde d}_\mu^*\ , \ {\tilde d}_\mu\ ]\ .
\eeq
In (I), we proved that ${\wtilde \Lambda}_{\pm,0}$ obey the $su(2)$-algebra and, further, they commute with any of the present $su(M+1)$-generators:
\beq
& &\left[\ {\wtilde \Lambda}_+\ , \ {\wtilde \Lambda}_-\ \right]=2{\wtilde \Lambda}_0\ , \qquad
\left[\ {\wtilde \Lambda}_0\ , \ {\wtilde \Lambda}_{\pm}\ \right]=\pm{\wtilde \Lambda}_{\pm}\ , 
\label{7}\\
& &\left[\ {\wtilde \Lambda}_{\pm,0}\ , \ {\rm any\ of\ }({\wtilde S}^i,\ {\wtilde S}_i,\ {\wtilde S}_j^i)\ \right]=0\ . 
\label{8}
\eeq
In the relation (\ref{5}) with (\ref{3}), we can see the following about a series of the single-particle states $(0,\mu), \ (1,\mu), \cdots ,\ (M,\mu)$ 
appearing in the states under consideration: 
The operation of ${\wtilde \Lambda}_+$ can work only in the case where this series is fully vacant and in any other case, 
this series vanishes. 
On the other hand, if this series is fully occupied, 
the operation of ${\wtilde \Lambda}_-$ does not make this series vanish and for any other case, this series vanishes.

In (I) and, of course, \citen{6}, we called the set $({\wtilde \Lambda}_{\pm,0})$ as the auxiliary $su(2)$-algebra in the $su(M+1)$-Lipkin model. 
However, in (I), the explicit form of ${\wtilde \Lambda}_0$ is presented in two simple cases with $M=1$ and 2. 
We can show that ${\wtilde \Lambda}_0$ is generally expressed as 
\beq
& &{\wtilde \Lambda}_0=\frac{1}{2}\sum_{\mu=1}^{2\Omega}\left(\prod_{i=0}^{M}{\tilde \nu}_{i,\mu}-\prod_{i=0}^{M}(1-{\tilde \nu}_{i,\mu})\right)\ , 
\label{9}\\
& &{\tilde \nu}_{i,\mu}={\tilde c}_{i,\mu}^*{\tilde c}_{i,\mu}\ . 
\label{10}
\eeq
Hereafter, we will use the following operators: 
\beq\label{11}
{\tilde \nu}_i=\sum_{\mu=1}^{2\Omega}{\tilde \nu}_{i,\mu}\ , \qquad
{\wtilde N}=\sum_{i=0}^M{\tilde \nu}_i \ . 
\eeq
Here, ${\tilde \nu}_i$ and ${\wtilde N}$ denote the fermion number operator in the level $i$ and the total, respectively. 
The case with $M=1$ is given as 
\beq\label{12}
{\wtilde \Lambda}_+=\sum_{\mu=1}^{2\Omega}{\tilde c}_{0,\mu}^*{\tilde c}_{1,\mu}^*\ , \qquad
{\wtilde \Lambda}_-=\sum_{\mu=1}^{2\Omega}{\tilde c}_{1,\mu}{\tilde c}_{0,\mu}\ , \qquad
{\wtilde \Lambda}_0=\frac{1}{2}{\wtilde N}-\Omega\ . 
\eeq
The form (\ref{12}) has been reported in \citen{6}. 
Since $[{\wtilde \Lambda}_0\ ,\ {\wtilde \Lambda}_{\pm}]=\pm{\wtilde \Lambda}_{\pm}$ and 
$[{\wtilde N}\ , \ {\wtilde \Lambda}_{\pm}]=\pm(M+1){\wtilde \Lambda}_{\pm}$, ${\wtilde \Lambda}_0$ can be expressed  
in the form 
\beq\label{13}
{\wtilde \Lambda}_0=\frac{1}{M+1}{\wtilde N}-{\wtilde {\cal L}}\ , \qquad
\left[\ {\wtilde {\cal L}}\ , \ {\wtilde \Lambda}_{\pm}\ \right]=0\ . 
\eeq
Of course, ${\wtilde {\cal L}}$ satisfies 
\beq\label{14}
\left[\ {\wtilde {\cal L}}\ , \ {\rm any\ of}\ \left({\wtilde S}^i,\ {\wtilde S}_i,\ {\wtilde S}_j^i\right)\ \right]=0\ .
\eeq
On the basis of the above relations, we will discuss how to construct the minimum weight state of the $su(M+1)$-Lipkin model in arbitrary fermion number. 
As was mentioned in the introductory part, the argument in this paper may be 
supplement to that given in (I).

Let $\kket{m}$ denote the minimum weight state. 
It obeys the following condition:
\bsub\label{15}
\beq
& &{\wtilde S}_i\kket{m}={\wtilde S}_i^j\kket{m}=0\ , \qquad (i>j)
\label{15a}\\
& &{\wtilde S}_i^i\kket{m}=-\sigma_i\kket{m}\ . 
\label{15b}
\eeq
\esub
The condition (\ref{15b}) can be rewritten in the form 
\beq\label{16}
({\tilde \nu}_i-{\tilde \nu}_0)\kket{m}=-\sigma_i\kket{m}\ . \qquad
(i=1,\ 2,\cdots ,\ M)
\eeq
For the state $\kket{m}$ obeying the condition (\ref{15}), we have the relation 
\bsub\label{17}
\beq
& &{\tilde \nu}_{k,\mu}{\wtilde S}_i\kket{m}={\wtilde S}_i{\tilde \nu}_{k,\mu}\kket{m}=0\ , 
\label{17a}\\
& &{\tilde \nu}_{k,\mu}{\wtilde S}_i^j\kket{m}={\wtilde S}_i^j{\tilde \nu}_{k,\mu}\kket{m}=0\ ,\qquad (i>j) 
\label{17b}\\
& &{\tilde \nu}_{k,\mu}{\wtilde S}_i^i\kket{m}={\wtilde S}_i^i{\tilde \nu}_{k,\mu}\kket{m}=-\sigma_i{\tilde \nu}_{k,\mu}\kket{m}\ . 
\label{17c}
\eeq
\esub
Then, as a possible choice, it may be permitted to set up the relation 
\beq\label{18}
{\tilde \nu}_{k,\mu}\kket{m}=\nu_{k,\mu}\kket{m}\ , \qquad 
\nu_{k,\mu}=1\ \ {\rm or}\ \ 0\ .
\eeq
If $\nu_{k,\mu}=1$ and 0, the single-particle state $(k,\mu)$ is occupied and vacant, respectively, for the fermion. 
As was already mentioned, one-fermion transfer through the $su(M+1)$-generators (\ref{1}) occurs between $(i,\mu)$ and $(j,\mu)$ with 
$i\neq j$ including $i$ or $j=0$. 
It should be noted that $\mu$ does not change. 
Then, the condition (\ref{15a}) tells us that in the state $\kket{m}$, the one-fermion 
transfer occurs in the case from the state $(i,\mu)$ to the lower state $(j,\mu)$, i.e., 
$i>j$. 
%This can be interpreted as follows: 
If the state $(j,\mu)$ is occupied, i.e., $\nu_{j,\mu}=1$, then, by the Pauli principle, this transfer is forbidden. 
Therefore, in the state $\kket{m}$, as the single-particle level becomes higher, i.e.,$i$ increases, the occupation number of the fermions in the state $i$ decreases. 
In our present model. we cannot find any condition which interferes with the relation (\ref{18}).

For searching the state $\kket{m}$, we introduce the state $\kket{m_0}$ governed by the conditions 
\beq
& &{\wtilde S}_i\kket{m_0}={\wtilde S}_i^j\kket{m_0}=0\ , \quad (i>j),\qquad
{\wtilde S}_i^i\kket{m_0}=-\sigma_i\kket{m_0}\ , 
\label{19}\\
& &{\wtilde \Lambda}_-\kket{m_0}=0\ , \qquad {\wtilde \Lambda}_0\kket{m_0}=-\lambda\kket{m_0}\ . 
\label{20}
\eeq
The condition (\ref{19}) is identical to the relation (\ref{15}). 
The relation (\ref{8}) teaches us that the relations (\ref{19}) and (\ref{20}) are compatible with each other. 
If $\kket{m_0}$ is obtained, $\kket{m}$ can be expressed in the form 
\beq\label{21}
\kket{m}=\left({\wtilde \Lambda}_+\right)^{\lambda+\lambda_0}\kket{m_0}\ . \qquad
(\lambda_0=-\lambda,\ -\lambda+1, \cdots ,\ \lambda-1,\ \lambda)
\eeq
As was already mentioned, if in $\kket{m_0}$, the series of the single-particle states\break
$(0,\mu),\ (1,\mu),\cdots ,\ (M,\mu)$ is fully vacant, 
we have ${\wtilde \Lambda}_+\kket{m_0}\neq 0$ and in any other case, 
${\wtilde \Lambda}_+\kket{m_0}=0$. 
If in $\kket{m_0}$, this series is fully occupied, we obtain ${\wtilde \Lambda}_-\kket{m_0}\neq 0$ and in any other case, ${\wtilde \Lambda}_-\kket{m_0}=0$.

%The, for searching $\kket{m}$, it may be enough to obtain $\kket{m_0}$. 
%For a moment, we forget the condition (\ref{19}). 
%If, in $\kket{m_0}$, all of the single-particle states $(i,1)$, $(i,2),\cdots ,\ (i,2\Omega)$ are vacant, we have 
%${\wtilde \Lambda}_-\kket{m_0}=0$. 
%On the other hand, if in $\kket{m_0}$, all of the single-particle state $(0,\mu)$, $(1,\mu),\cdots ,\ (M,\mu)$ are vacant, 
%we have ${\wtilde \Lambda}_+\kket{m_0}\neq 0$. 
%Here, it should be noted that the quantum numbers $i$ and $\mu$ are unchanged, respectively. 

\begin{figure}[t]
\begin{center}
\includegraphics[height=7.5cm]{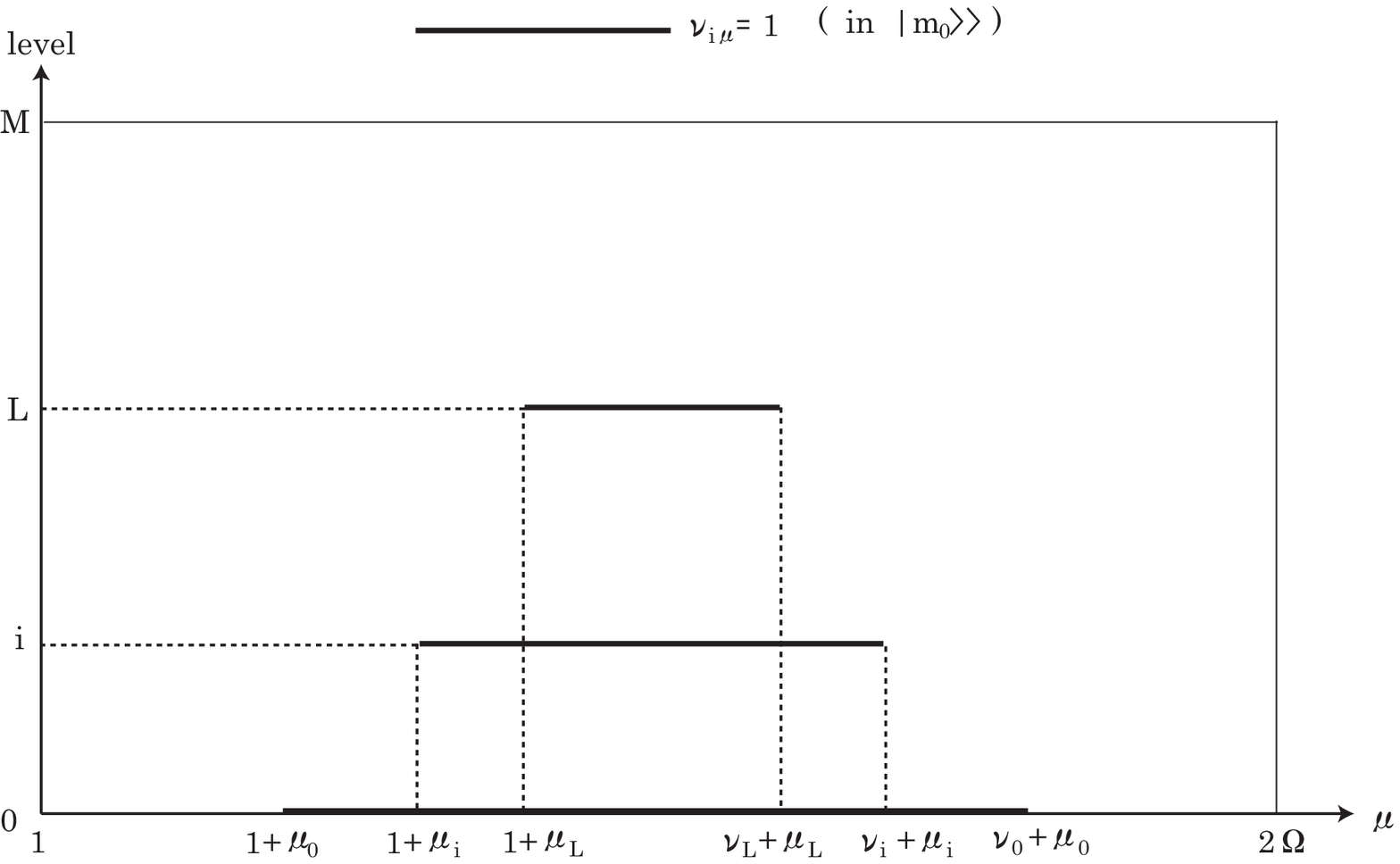}
\caption{The schematic level scheme and occupation numbers are illustrated.
}
\label{fig:1}
\end{center}
\end{figure}

On the basis of the above consideration, first, we construct the simplest example of $\kket{m_0}$. 
We treat the following case: 
For the level $i\ (=0,\ 1,\ 2,\cdots ,\ L)$, $\nu_{i,\mu}=1$ in the range $\mu=1+\mu_i,\ 2+\mu_i,\cdots ,\ \nu_{i-1}+\mu_i,\ \nu_i+\mu_i$ 
and in the remaining ranges $\nu_{i,\mu}=0$. 
The above-mentioned scheme is illustrated in Fig.\ref{fig:1}, which teaches us that $\nu_i$ fermions occupy the level $i$. 
Any fermion does not occupy the levels $i=L+1,\ L+2,\cdots ,\ M$. 
The consideration on the one-fermion transfer and the operation of ${\wtilde \Lambda}_{\pm}$ gives us the following relation : 
\bsub\label{21sub}
\beq
& &
\nu_0 \geq \nu_1 \geq \cdots \geq \nu_L>0\ , \qquad
\nu_{L+1}=\nu_{L+2}=\cdots =\nu_{M}=0\ , 
\label{21a}\\
& &0\leq \mu_0 \leq \mu_1 \leq \cdots \leq \mu_L\ , 
\label{21b}\\
& &
\nu_0+\mu_0\leq 2\Omega\ . 
\label{21c}
\eeq
\esub
Thus, $\kket{m_0}$ can be expressed in the form 
\beq\label{22}
\kket{m_0}=\prod_{i=0}^{L}\prod_{\mu=1+\mu_i}^{\nu_i+\mu_i}{\tilde c}_{i,\mu}^*\kket{0}\ . \qquad
\left({\tilde c}_{i,\mu}\kket{0}=0\right)
\eeq
Of course, $\{\ \kket{m_0}\ \}$ forms the normalized orthogonal set. 
%
%%%%%%%%%%%%%%%%%%%%%%%%%%%%%%%%%%%%%%%%%%%%%%%%%%%%%%%%%%%%%%%%%%%%%%
%%%%%%%%%%%%%%%%%%%%%%%%%%%%%%%%%%%%%%%%%%%%%%%%%%%%%%%%%%%%%%%%%%%%%%%%
%
The quantity $\sigma_i$ given in the relation (\ref{19}) is expressed as 
\beq\label{23}
\sigma_i=\left\{
\begin{array}{ll}
\displaystyle \nu_0-\nu_i\ , & (i=1,\ 2, \cdots ,\ L)\\
\displaystyle \nu_0\ , & (i=L+1,\ L+2, \cdots ,\ M)\\
\end{array}
\right.
\eeq
The above relations lead us to 
\beq\label{24}
\sigma_1 \leq \sigma_2 \leq \cdots \leq \sigma_L < \sigma_{L+1}=\sigma_{L+2}=\cdots =\sigma_M\ (=\nu_0)\ . 
\eeq
Hereafter, instead of $\{\sigma_i\}$, we formulate the minimum weight states by $\{\nu_i\}$. 
With the use of the relation (\ref{9}) for ${\wtilde \Lambda}_0$, we obtain $\lambda$ in the form 
\beq\label{25}
\lambda=\frac{1}{2}\left(\sum_{\mu=1}^{\mu_0}1+\sum_{\mu=\nu_0+\mu_0+1}^{2\Omega}1\right)=\Omega-\frac{1}{2}\nu_0\ . 
\eeq
%
%%%%%%%%%%%%%%%%%%%%%%%%%%%%%%%%%%%%%%%%%%%%%%%%%%%%%%%%%%%%%%%%%%%%%%
\begin{figure}[t]
\begin{center}
\includegraphics[height=8.5cm]{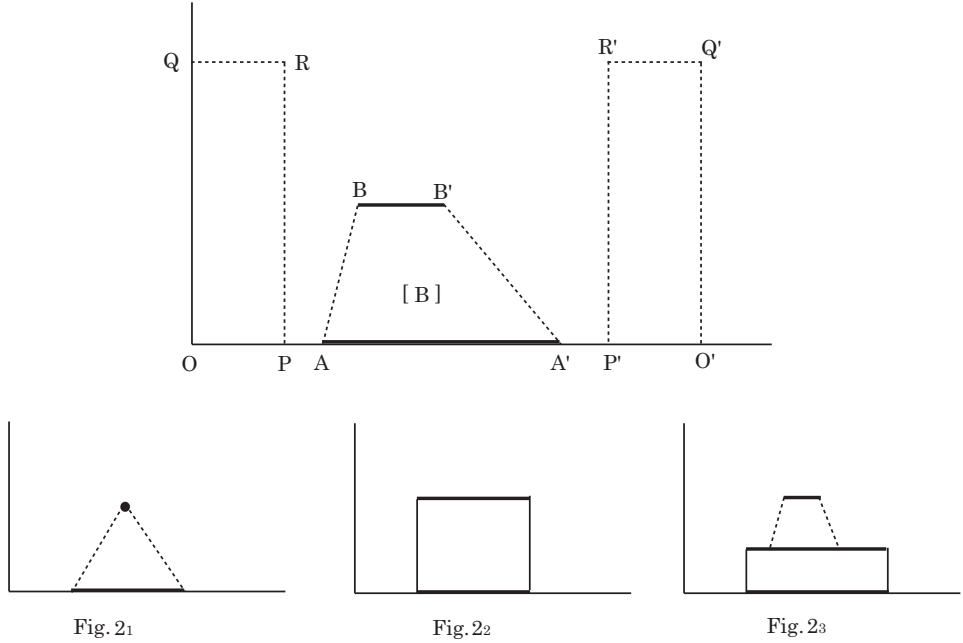}
\caption{Schematic feature of Fig.\ref{fig:1} is depicted.
}
\label{fig:2}
\end{center}
\end{figure}
%%%%%%%%%%%%%%%%%%%%%%%%%%%%%%%%%%%%%%%%%%%%%%%%%%%%%%%%%%%%%%%%%%%%%%%%
%
The total fermion number for $\kket{m_0}$, which is denoted as $N_{\rm min}$, is given in the form 
\beq\label{26}
& &N_{\rm min}=\sum_{i=0}^{L}\nu_i=\nu_0+\nu(L)\ , \nonumber\\
& &\nu(L)=\sum_{i=1}^L \nu_i\ . 
%(L+1)\nu_0-\sum_{i=1}^{L}\sigma_i
%=(M+1)\nu_0-\sum_{i=1}^{M}\sigma_i\ . 
\eeq 
%Here, we used the relation (\ref{23}). 
Then, ${\cal L}$, 
%with the use of the relation (\ref{13}), 
the eigenvalue of ${\wtilde {\cal L}}$ which is defined in the relation (\ref{13}), is given by 
\beq\label{27}
{\cal L}=\Omega-\frac{M-1}{2(M+1)}\nu_0+\frac{1}{M+1}\nu(L)\ . 
\eeq
The quantity $\lambda_0$ for any fermion number, $N$, is expressed as 
\beq\label{28}
\lambda_0=\frac{1}{M+1}N-\left(\Omega-\frac{M-1}{2(M+1)}\nu_0+\frac{1}{M+1}\nu(L)\right)\ . 
\eeq
Of course, $\lambda_0$ obey the inequality 
\beq\label{29}
-\lambda \leq\lambda_0 \leq \lambda\ . 
\eeq
The case with $\lambda_0=\lambda$ gives us the maximum fermion number, $N_{\rm max}$. 
With use of the relations (\ref{26}) and (\ref{28}) for $\lambda_0=\lambda$, we have 
\beq\label{30}
N_{\rm max}&=&
2\Omega\cdot (M+1)-M\nu_0-\nu(L)\nonumber\\
&=&N_{\rm min}+(M+1)\cdot 2\lambda\ . 
\eeq
Through the relation (\ref{21}), the normalized state $\kket{m}$ is given as 
\beq\label{33}
\kket{m}=\sqrt{\frac{(\lambda-\lambda_0)!}{(2\lambda)!(\lambda+\lambda_0)!}}\left({\wtilde \Lambda}_+\right)^{\lambda+\lambda_0}\kket{m_0}\ . 
\eeq
Therefore, we do not have the trouble to the normalization.

\begin{figure}[t]
\begin{center}
\includegraphics[height=4.5cm]{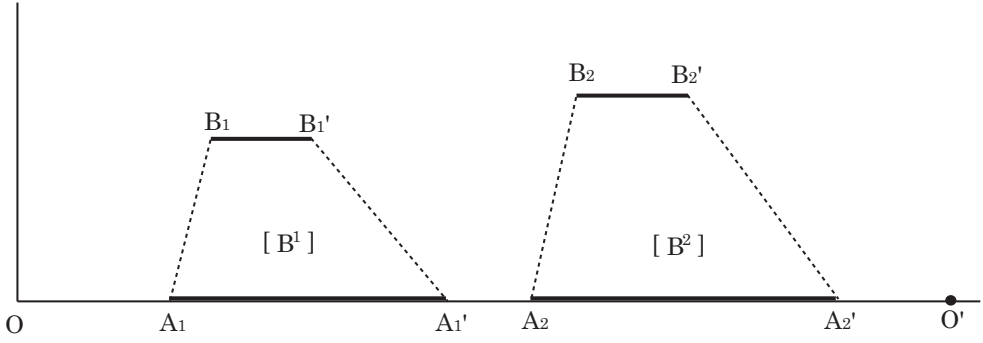}
\caption{The case composed of two blocks is depicted schematically.
}
\label{fig:3}
\end{center}
\end{figure}

Schematic feature of Fig.\ref{fig:1} is depicted in Fig.2. 
In the upper figure in Fig.\ref{fig:2}, the block [B] surrounded by the points A, A$'$, B and B$'$ is trapezoid-like in shape. 
The segments AA$'$ and BB$'$ are parallel with each other and the sections AB and A$'$B$'$ are stepwise. 
If ${\overline {\rm BB'}}=0$ and ${\overline {\rm BB'}}={\overline{\rm AA'}}$, [B] becomes triangle-like (Fig.2$_1$) and the rectangular (Fig.2$_2$), respectively. 
Further, as a possible shape of [B], we have the figure obtained by piling the trapezoid-like figure on the rectangle (Fig.2$_3$). 
The number of the lattice points in [B] corresponds to the total fermion number in $\kket{m_0}$, $N_{\rm min}$. 
Intervals OP and P$'$O$'$ are related with the number of times of the operation of ${\wtilde \Lambda}_+$. 
If ${\overline{\rm OA}}\ ({\overline{\rm A'O'}})=0$, the point P (P$'$) disappears and the operation of ${\wtilde \Lambda}_+$ is meaningless. 
If ${\overline {\rm OA}}\ ({\overline{\rm A'O}})\geq 1$, the interval OP (P$'$O$'$) for the operation of ${\wtilde \Lambda}_+$ becomes meaningful. 
Next, we consider the case composed of two blocks [B$^1$] and [B$^2$], which is depicted in Fig.\ref{fig:3}. 
The discussion in the case with [B] can be applied to the sides OA$_1$ and A$'_2$O$'$ in the present case. 
Then, it may be enough to discuss the interrelation between the points A$'_1$ and A$_2$. 
It is easily verified that if ${\overline{{\rm A'}_1{\rm A}_2}}\leq 1$, the interval A$'_1$A$_2$ does not contributes to the operation of ${\wtilde \Lambda}_+$,
but if ${\overline {{\rm A'}_1{\rm A}_2}}>1$, A$'_1$A$_2$ contributes to the operation of ${\wtilde \Lambda}_+$.

%
%%%%%%%%%%%%%%%%%%%%%%%%%%%%%%%%%%%%%%%%%%%%%%%%%%%%%%%%%%%%%%%%%%%%%%
%%%%%%%%%%%%%%%%%%%%%%%%%%%%%%%%%%%%%%%%%%%%%%%%%%%%%%%%%%%%%%%%%%%%%%%%
%

%
%%%%%%%%%%%%%%%%%%%%%%%%%%%%%%%%%%%%%%%%%%%%%%%%%%%%%%%%%%%%%%%%%%%%%%
\begin{figure}[b]
\begin{center}
\includegraphics[height=5cm]{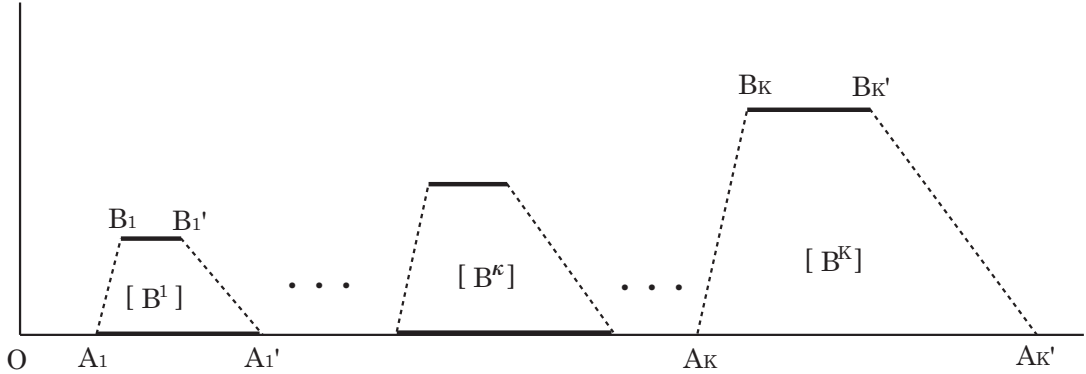}
\caption{The case composed of $K$ blocks is depicted schematically.
}
\label{fig:4}
\end{center}
\end{figure}
%%%%%%%%%%%%%%%%%%%%%%%%%%%%%%%%%%%%%%%%%%%%%%%%%%%%%%%%%%%%%%%%%%%%%%%%
%

If we follow the above consideration, it may be easy to treat the general case, i.e., the case with $K$ blocks. 
First, we prepare the blocks labeled by $\kappa =1,\ 2,\cdots ,\ K$ and [F$^\kappa$] denotes the $\kappa$-th block. 
Then, it is enough to line them up along the $\mu$-axis (Fig.\ref{fig:4}). 
Of course, it is natural to avoid overlapping with each other. 
The relation (\ref{26}) is useful in the present case and, then, we can construct $\kket{m_0}$. 
The relation (\ref{33}) gives us $\kket{m}$. 
The number of the lattice points $N^{\kappa}$ in [B$^{\kappa}$] and $N_{\rm min}\ (=\sum_{\kappa=1}^K N^{\kappa})$ are given by 
\beq
%N^\kappa=\nu_0^{\kappa}+\sum_{i=1}^{L^{K}}\nu_i^{\kappa}\ .
& &N_{\rm min}=\nu_0+\nu(L)\ , 
\label{34}\\
%Then, $N_{\rm min}$ is expressed in the form 
%\beq
%& &N_{\rm min}=\sum_{\kappa=1}^K N^{\kappa}
%=\nu_0+\sum_{i=1}^{L}\nu_i\ , 
%\label{35}\\
& &\nu_0=\sum_{\kappa=1}^K \nu_0^{\kappa}\ , \qquad 
\nu_i=\sum_{\kappa=1}^K \nu_i^{\kappa}\ ,\qquad
\nu(L)=\sum_{i=1}^L\nu_i\ , 
\label{35}\\
& &L={\rm max}(L^1,\cdots , L^{K})\ .  
\label{36}
\eeq

%
%%%%%%%%%%%%%%%%%%%%%%%%%%%%%%%%%%%%%%%%%%%%%%%%%%%%%%%%%%%%%%%%%%%%%%
\begin{figure}[t]
\begin{center}
\includegraphics[height=11cm]{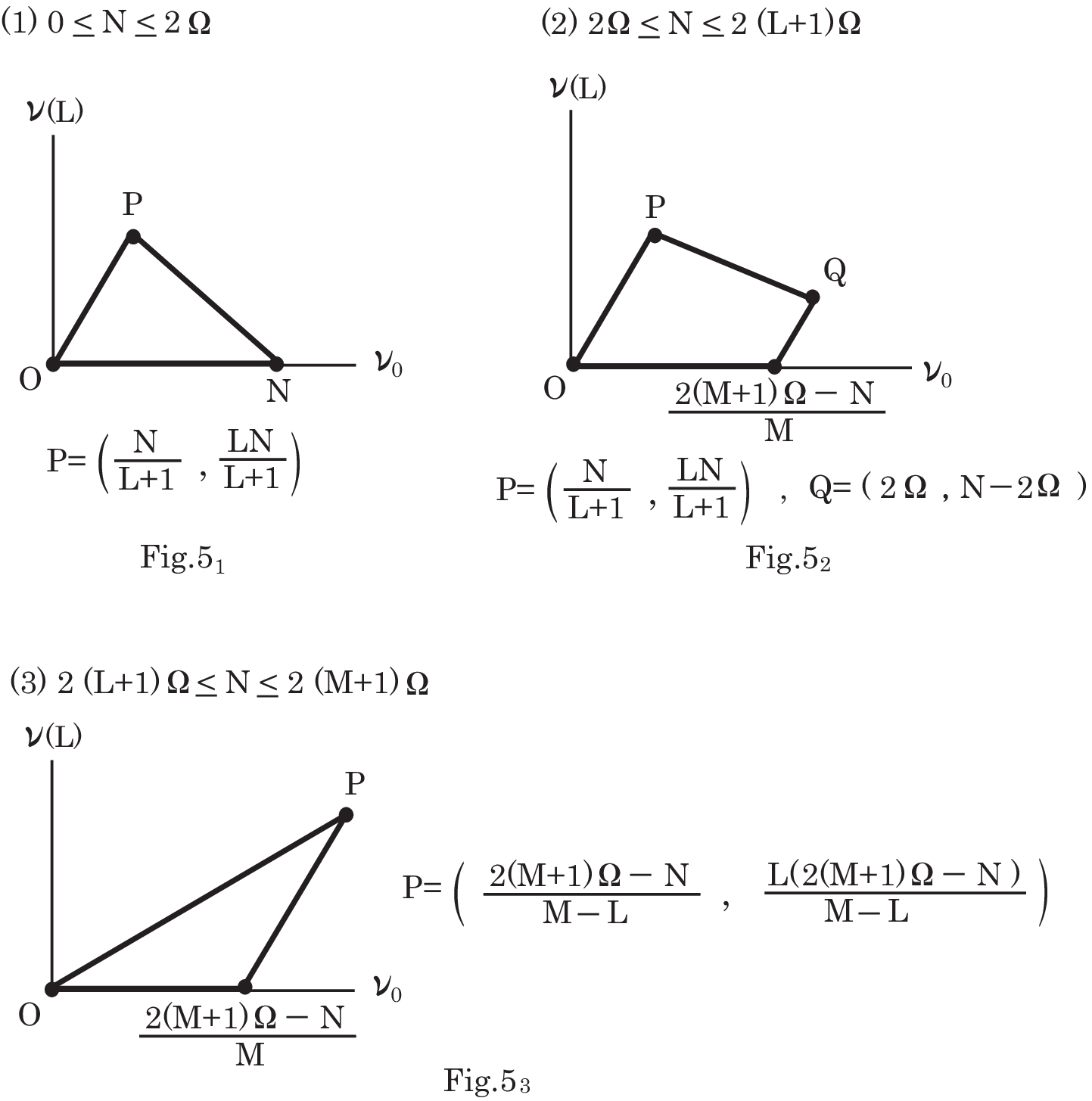}
\caption{The domains in which the inequalities (\ref{21a}) and (\ref{29}) are realized are shown 
as the area inside the triangle and quadrilateral, respectively.
}
\label{fig:5}
\end{center}
\end{figure}
%%%%%%%%%%%%%%%%%%%%%%%%%%%%%%%%%%%%%%%%%%%%%%%%%%%%%%%%%%%%%%%%%%%%%%%%
%

For the other relations, we have the same expressions as those given in the simplest example.

%As a final remark, 
We will show the domains on the plane $(\nu_0,\nu(L))$, where the inequalities (\ref{21a}) and (\ref{29}) are realized. 
These inequalities lead us to the following relations:
\bsub\label{36}
\beq
& &\nu(L)\leq L\nu_0\ , 
\label{36a}\\
& &\nu(L)\leq N-\nu_0\ , 
\label{36b}\\
& &\nu(L)\geq -(2(M+1)\Omega -N)+M\nu_0\ . 
\label{36c}
\eeq
\esub
The relation (\ref{36a}) is derived from the inequalities (\ref{21a}), 
but the reverse is not true. 
In the case with $L=0$, 
simply we can show the following domains:
\bsub\label{37}
\beq
& &({\rm i})\ \ 0\leq N \leq 2\Omega\ ,\qquad 
0\leq \nu_0 \leq N\ , \qquad \nu(0)=0\ , 
\label{37a}\\
& &({\rm ii})\ 
2\Omega \leq N \leq 2(M+1)\Omega \ , \qquad 
0\leq \nu_0 \leq \frac{2(M+1)\Omega-N}{M}\ , \qquad
\nu(0)=0 \ . \ \ \ \ 
\label{37b}
\eeq
\esub
In the case with $1\leq L \leq M-1$, the domains are depicted in Figs. $5_1$, $5_2$ and $5_3$.

\begin{figure}[t]
\begin{center}
\includegraphics[height=6.5cm]{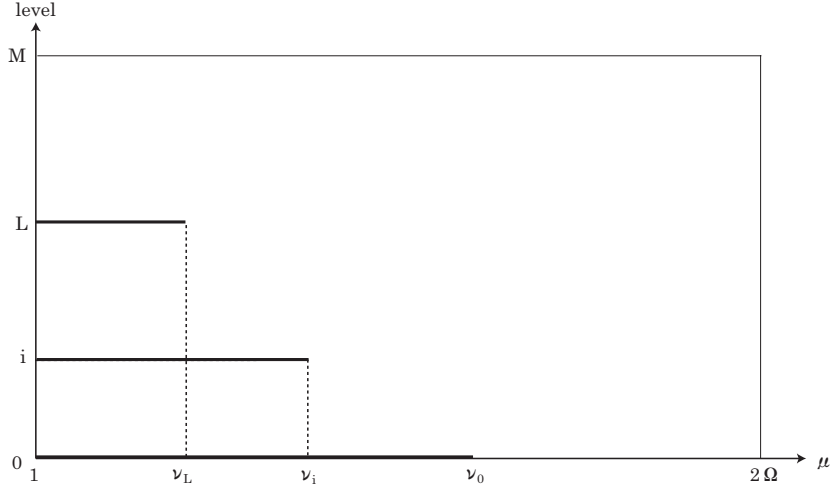}
\caption{The schematic level scheme and occupation numbers are illustrated.
}
\label{fig:6}
\end{center}
\end{figure}

Finally, we will give a remark. 
As was mentioned in the introductory part, the algebraic approach to many-body theories starts in the task 
how to express the minimum weight states. 
In this paper, we presented a practical scheme for constructing the minimum weight states in the space 
spanned by $i=0,\ 1,\cdots ,\ M$ and $\mu=1,\ 2,\cdots ,\ 2\Omega$. 
If we encounter the system which contains the components violating 
the $su(n)$-symmetry in a greater or less degree, we must treat plural minimum weight states simultaneously. 
In such situations, our scheme may be useful. 
On the other hand, we know the case where it may be enough to adopt a single minimum weight state. 
In this case, our scheme becomes much simpler. 
It may be permitted to change the numbering of $\mu$ appropriately. 
It suggests us to put $\mu_i=0$ $(i=0,\ 1,\cdots ,\ L)$ in Fig.\ref{fig:1} and 
it becomes Fig.\ref{fig:6}. 
In this case, $\kket{m_0}$ can be expressed in the form 
\beq\label{38}
\kket{m_0}=\prod_{i=0}^L\prod_{\mu=1}^{\nu_i}{\tilde c}_{i,\mu}^*\kket{0}\ . 
\eeq
The form (\ref{38}) is useful for the Hamiltonian expressed in terms of the generators (\ref{1}), 
for example, such as that given in the relations (I.2.6) and (I.2.9).

%
%%%%%%%%%%%%%%%%%%%%%%%%%%%%%%%%%%%%%%%%%%%%%%%%%%%%%%%%%%%%%%%%%%%%%%
%%%%%%%%%%%%%%%%%%%%%%%%%%%%%%%%%%%%%%%%%%%%%%%%%%%%%%%%%%%%%%%%%%%%%%%%
%

Thus, we were able to obtain the minimum weight states $\kket{m}$ in a practical scheme. 
Needless to say, the above treatment is predicted on the property of the one-fermion transfer proper to the 
generators in the $su(n)$-Lipkin model. 
In this transfer, the quantum number $\mu$ does not change. 
Further, we should stress that the auxiliary $su(2)$-algebra also plays a central role: 
With the aid of this algebra, $\kket{m}$ is derived from $\kket{m_0}$. 

In the forthcoming paper, we will propose an idea of the random phase approximation based on the minimum weight state (\ref{38}) 
and discuss the phase change observed under this approximation.

\section*{Acknowledgment}

Two of the authors (Y.T. and M.Y.) would like to express their thanks to 
Professor J. da Provid\^encia and Professor C. Provid\^encia, two of co-authors of this paper, 
for their warm hospitality during their visit to Coimbra in spring of 2015. 
The author, M.Y., would like to express his sincere thanks to 
Mrs M. Nakamura and
\break
Mrs H. Tani for their encouragement.


\begin{thebibliography}{9}


\bibitem{1}
H. J. Lipkin, N. Meshkov and A. Glick, Nucl. Phys. {\bf 62}, 188 (1965).


\bibitem{2}
S. Li, A. Klein and R. M. Dreizler, J. Math Phys. {\bf 11}, 975 (1970).\\
N. Meshkov, Phys. Rev. C {\bf 3}, 2214 (1971).\\
S. Okubo, J. Math. Phys. {\bf 16}, 528 (1975).\\
A. Klein, Nucl. Phys. A {\bf 347}, 3 (1980).
%A. Klein and E. R. Marshalek, Rev. Mod. Phys. {\bf 63}, 375 (1991).


\bibitem{3}
Y. Tsue, C. Provid\^encia, J. da Provid\^encia and M. Yamamura, 
Prog. Theor. Exp. Phys. {\bf 2016}, 083D03 (2016). 


\bibitem{4}
Y. Tsue, C. Provid\^encia, J. da Provid\^encia and M. Yamamura, 
Prog. Theor. Exp. Phys. {\bf 2016}, 083D04 (2016). 



\bibitem{5}
Y. Tsue, C. Provid\^encia, J. da Provid\^encia and M. Yamamura, 
to appera in Prog. Theor. Exp. Phys. {\bf 2017}. 
%, 083D04 (2016). 



\bibitem{6}
Y. Tsue, C. Provid\^encia, J. da Provid\^encia and M. Yamamura, 
Prog. Theor. Exp. Phys. {\bf 2015}, 063D01 (2015). 



\end{thebibliography}
\end{document}